\documentclass[aps,prb,showpacs,twocolumn,floats,epsfig]{revtex4}

\usepackage{amsmath}
\usepackage{epsfig}
\usepackage{graphicx}
\usepackage{amsfonts}
\usepackage[dvips]{color}

\newcommand{\ct}{\cite}
\newcommand{\bi}{\bibitem}
\newcommand{\be}{\begin{equation}}
\newcommand{\ee}{\end{equation}}
\newcommand{\ba}{\begin{eqnarray}}
\newcommand{\ea}{\end{eqnarray}}
\newcommand{\al}{\alpha}

\newcommand{\de}{\delta}
\newcommand{\la}{\lambda}
\newcommand{\si}{\sigma}
\newcommand{\dg}{\dagger}
\newcommand{\non}{\nonumber}

\begin{document}

\title{Quantum fidelity for one-dimensional Dirac fermions and 
two-dimensional Kitaev model in the thermodynamic limit}
\author{Victor Mukherjee${}^1$, Amit Dutta${}^2$ and Diptiman Sen${}^3$}
\affiliation{$^{1,2}$Department of Physics, Indian Institute of Technology
Kanpur 208 016, India \\
${}^3$Center for High Energy Physics, Indian Institute of Science, Bangalore
560 012, India}

\begin{abstract}
We study the scaling behavior of the fidelity ($F$) in the thermodynamic limit
using the examples of a system of Dirac fermions in one dimension and the 
Kitaev model on a honeycomb lattice. We show that the thermodynamic fidelity 
inside the gapless as well as gapped phases follow power-law scalings, with 
the power given by some of the critical exponents of the system. The generic 
scaling forms of $F$ for an anisotropic quantum critical point for both 
thermodynamic and non-thermodynamic limits have been derived and verified for 
the Kitaev model. The interesting scaling behavior of $F$ inside the gapless 
phase of the Kitaev model is also discussed.
Finally, we consider a rotation of each spin in the Kitaev model around the 
$z$ axis and calculate $F$ through the overlap between the ground states for 
angle of rotation $\eta$ and $\eta+d\eta$, respectively. We thereby show that 
the associated geometric phase vanishes. We have supplemented our analytical 
calculations with numerical simulations wherever necessary.
\end{abstract}


\maketitle

\section{Introduction}

A quantum phase transition 
\ct{chakrabarti96, sachdev99,continentino,sondhi97,vojta03} driven 
exclusively by quantum fluctuations at zero temperature is associated 
with a dramatic change in the symmetry of the ground state of a 
many-body quantum Hamiltonian. A number of measures from quantum information 
theory such as entanglement \ct{osterloh02, roncaglia06}, entanglement 
entropy \ct{vidal03,kitaev061}, Loschmidt echo \ct{song06}, decoherence 
\ct{damski10} and quantum discord \ct{olliver01,dillenschneider08} are able 
to capture the singularities associated with a quantum critical point (QCP).
Consequently, there is a recent upsurge in the investigation of quantum 
critical systems from the perspective of quantum information theory in an 
attempt to establish a bridge between these two fields 
\ct{amico08,latorre09,duttarmp10}.
 
An important information theoretic concept that is being investigated 
extensively for quantum critical systems is the quantum fidelity ($F$) 
\ct{zanardi06,venuti07,giorda07,zhou081,shigu08,gurev10,schwandt09,gritsev09,
yang08,lin09,grandi10,polkovnikovrmp,mukherjee10, mukherjee11,zhou082,zhou083,
zhao09,znidaric03,ma08,eriksson09,rams11}.
Let us consider a $d$-dimensional Hamiltonian $H(\la)$ which contains
an externally tunable parameter $\la$, such that the system is at a
QCP when $\la=0$. Considering two ground state wave functions
$|\psi_0(\la)\rangle$ and $|\psi_0 (\la + \de) \rangle$, 
which are infinitesimally separated in the parameter space as $\de \to 0$, 
we define the fidelity as
\ba F(\la, \la + \de) &=& |\langle \psi_0(\la)| \psi_0 (\la + \de)\rangle| 
\non \\
&=& 1 - \frac{\de^2}{2}L^d\chi_F + \cdots, \label{fidelity_exp} \ea
where $L$ is the linear dimension of the system, and $\de$ denotes a small 
change in the parameter $\la$. The first non-vanishing term in the 
expansion of $F$, namely, the fidelity susceptibility $\chi_F$, provides a 
quantitative measure of the rate of change of the ground state under an
infinitesimal variation of $\la$. By exhibiting a sharp decay around 
a QCP even for a finite size system, the fidelity turns out to be one of the 
fundamental probes for detecting the ground state singularities associated 
with a quantum phase transition without making reference to an order parameter.
At the same time, the fidelity susceptibility defined through the 
relation $\chi_F = -(2/L^d) (\ln F)/\de^2|_{\de \to 0}= -(1/L^d) 
(\partial^2 F/\partial \de^2)$ usually diverges with the system size in
a universal power-law fashion with an exponent given in terms of some of
the quantum critical exponents. For a marginal or relevant perturbation 
$\la$, the application of the adiabatic perturbation theory leads to a 
generic scaling form \ct{venuti07, gurev10, gritsev09, schwandt09, grandi10} 
of $\chi_F$ given by $\chi_F (\la = 0) \sim L^{2/\nu - d}$ at the QCP, 
whereas away from the QCP, the scaling changes to $\chi_F \sim 
|\la|^{\nu d - 2}$ for $L > \la^{-\nu}$; here $\nu$ is the critical
exponent describing the divergence of the spatial correlation length close 
to the QCP, i.e., $\xi \sim \la^{-\nu}$. 

While the fidelity susceptibility approach usually assumes small $L$ and 
$\de \to 0$, the fidelity per site\ct{zhou082,zhou083,zhao09} has
been calculated in the thermodynamic limit for an arbitrary value of $\de$
in some recent studies \ct{zhou082,zhou083,zhao09}. 
In such cases, the fidelity differs significantly from unity unlike in 
Eq.~(\ref{fidelity_exp}). The fidelity per site is also able to indicate the 
appearance of a quantum phase transition in the spin-1/2 $XY$ model in a
transverse magnetic field. Another
measure of fidelity applied to mixed states, namely, the reduced fidelity
\ct{znidaric03,ma08,eriksson09} has also provided important insights into
quantum critical phenomena. We note that similar studies have been carried 
out on the scaling of the geometric phase \ct{pancharatnam56,berry84} which is 
closely related to the fidelity susceptibility\ct{venuti07} close to critical
\ct{carollo05,zhu06,hamma06} and multicritical \ct{patra11} points.

As seen from Eq.~(\ref{fidelity_exp}) and also the preceding discussion, the 
knowledge of $\chi_F$ should be sufficient to draw conclusions about the 
behavior of $F$ and the associated QCP for small system sizes and in the limit
$\de \to 0$ ($\de^2 L^d \chi_F/2 \ll 1$). However, in the thermodynamic 
limit ($L \to \infty$ at fixed $\de$), the expansion in 
Eq.~(\ref{fidelity_exp}) up to the lowest order becomes insufficient. 
Recently, Rams and Damski \ct{rams11} have proposed a generic scaling relation 
\ct{rams11} valid in the thermodynamic limit given by
\ba \ln{F(\la - \de, \la + \de)} \simeq -L^d|\de|^{\nu d} A
\left(\frac{\la}{|\de|}\right), \label{fidelity_generic} \ea
where $A$ is a scaling function; this relation interpolates between the 
fidelity susceptibility approach and the fidelity per site approach. In 
deriving the scaling relation in Eq.~(\ref{fidelity_generic}), it is assumed 
that the fidelity per site is well behaved in the limit $L \to \infty$, the 
QCP is determined by a single set of critical exponents, and $\nu d >2$ so 
that non-universal corrections are subleading \ct{gritsev09}. In particular, 
at the critical point $\la=0$, the fidelity, measured between the ground 
states at $+\de$ and $-\de$, is non-analytic in $\de$ and satisfies
the scaling $\ln{F} \sim -L^d |\de|^{\nu d}$. On the other hand, away from 
the QCP, i.e., for $|\de| \ll |\la| \ll 1$, the scaling gets modified
to $\ln{F} \sim -L^d \de^2 |\la|^{\nu d - 2}$. This scaling has been 
verified for an isolated quantum critical point using one-dimensional 
transverse Ising and $XY$ Hamiltonians \ct{rams11}. Moreover, near a QCP a 
cross-over has been observed from the thermodynamic limit ($L |\de|^{\nu} 
\gg 1 $) to the non-thermodynamic (small system) limit ($L |\de|^{\nu} \ll 1$) 
where the concept of fidelity susceptibility becomes useful. We note that 
Eq.~(\ref{fidelity_generic}) is an example of the Anderson orthogonality 
catastrophe \ct{anderson67} which states that the overlap of two states 
vanishes in the thermodynamic limit irrespective of their proximity to a QCP.
 
In this paper we investigate the scaling of the thermodynamic fidelity in a 
one-dimensional system of Dirac fermions with a mass perturbation
\ct{mahan00,gogolin98,vondelft98,giamarchi04,giuliani05} and the 
two-dimensional Kitaev model on a honeycomb lattice \ct{kitaev06,chen07} close
to or inside the gapless phases of their phase diagrams, thereby extending 
previous studies to more generic situations. For the one-dimensional system, 
we have verified the scaling predicted in Ref. \onlinecite {rams11}. We also 
propose a generic scaling form for the thermodynamic fidelity in the vicinity 
of an anisotropic quantum critical point (AQCP), and we verify it for the 
AQCP present in the Kitaev model phase diagram \ct{hikichi10}.

The paper is organized in the following way. In Sec. II, we analytically
derive the scaling relations of the thermodynamic fidelity for non-interacting
spinless massive Dirac fermions in one dimension and propose a generalization 
to the case of interacting fermions (called a Tomonaga-Luttinger liquid). In 
Sec. III, we concentrate on the fidelity of the Kitaev model on the hexagonal 
lattice for different values of the coupling parameters. We derive the
scaling laws for both the thermodynamic and non-thermodynamic limits
for an AQCP and verify these numerically for the Kitaev model. In Sec. IV, we 
calculate the overlap between two ground states of the Kitaev model; in one 
state, all the spins are rotated about the $z$ axis by an angle $\eta$, while 
in the other, they are rotated by an angle $\eta + d\eta$. We thus derive 
the form of the fidelity through an expansion in powers of $d\eta$.

\section{Dirac fermions in one dimension}

In this section we consider a system of spinless Dirac 
fermions in one dimension with a mass perturbation and verify the scaling of 
the thermodynamic fidelity as predicted in Ref. \onlinecite{rams11}. 
Let us first consider non-interacting fermions. The Hamiltonian we consider is
\ba H = \sum^{\infty}_{k > 0} {\left[k \left(c^{\dg}_k c_k - c^{\dg}_{-k}
c_{-k}\right) + m\left( c^{\dg}_k c_{-k} + c^{\dg}_{-k}c_k\right)\right]}, 
\label{ham1} \ea
where $c_k^{\dg}$ ($c_k$) is the fermionic creation (annihilation) operator
for wave vector $k$, $m$ is the mass, and we have set the velocity 
$v = 1$ for convenience. In the two-level system given by $c^{\dg}_k c_k 
+ c^{\dg}_{-k}c_{-k} = 1$, the Hamiltonian takes the form
\ba H &=& \sum^{\infty}_{k > 0}\left( \begin{array}{cc} c^{\dg}_k & 
c^{\dg}_{-k} \end{array} \right)h_k \left(\begin{array}{c} 
c_k \\ c_{-k}\end{array}\right), \non \\
\text{where}~~ h_k &=& \left(\begin{array}{cc} 
k & m \\ m & -k \end{array}\right). \ea
The normalized ground state of this is given by
\ba &&\psi(k,m) \non \\ 
&=& \frac{1}{\sqrt{2\left(k^2 + m^2 \right) + 2k\sqrt{k^2 + m^2}}}\left(
\begin{array}{c} m \\ -\sqrt{k^2 + m^2} - k \end{array}\right). \non \\
&& \ea
with the energy $E_k = - \sqrt{k^2 + m^2}$.

If we now consider two systems with masses $m_1$ and $m_2$, the fidelity 
between the two ground states is given by
\be F(m_1, m_2) ~=~ \prod_{k > 0} ~|\langle \psi (k,m_1)|\psi(k,m_2) \rangle |.
\label{fidelity_luttinger1} \ee
Note that we have taken $k$ to be strictly positive in 
all the equations above. It is important to exclude the mode with $k=0$, 
otherwise the fidelity is exactly equal to zero if $m_1$ and $m_2$ have
opposite signs; this is because $\langle \psi (0, m_1) | \psi (0, m_2) 
\rangle = 0$ if $m_1 m_2 < 0$. The simplest way to exclude a zero momentum
mode is to impose antiperiodic boundary conditions, $\psi (x=L) =-\psi (x=0)$,
so that $k_n = (\pi/L) (2n+1)$, where $L$ is the system size and $n=0,1,2,
\cdots$. (Note that the spacing between successive values of $k$ is $2\pi /
L$). We can then write Eq. (\ref{fidelity_luttinger1}) as 
\be F(m_1, m_2) ~=~ \prod_{n = 0}^\infty ~|\langle \psi (k_n,m_1)| 
\psi(k_n,m_2) \rangle |. \label{fidelity_luttinger2} \ee

Now we consider the case with $m_1 = m$ and $m_2 = -m$ so that the states lie 
on the two sides of the gapless critical point, and $m$ plays the role of
$\de$ discussed in the previous section. We find that 
\be \langle \psi (k_n, m )|\psi(k_n, - m )\rangle = \frac{k_n^2 + k_n 
\sqrt{k_n^2 +m^2}}{ k_n^2 + m^2 + k_n \sqrt{k_n^2 +m^2}}. \ee
Since $k_n = (\pi/L) (2n+1)$, we see that the fidelity is a function of a 
single parameter given by $mL$. We now consider two cases: (i) $mL \gg 1$ and 
(ii) $mL \ll 1$. In both cases, we will assume that $L \gg 1$. (Cases (i) and 
(ii) will be respectively called the thermodynamic and non-thermodynamic 
limits in the next section). 

In case (i), we can take $k$ to be a continuous variable so that the fidelity
is given by an integral, 
\be F(m, -m) = \exp\left[L \int^{\infty}_0 {\frac{dk}{2\pi} \ln| \langle 
\psi(k,m_1)| \psi(k,m_2) \rangle |} \right], \label{fidelity_luttinger3} \ee
By writing $k = mx$ in the integral, we find that $F(m, -m)$ is given by 
$e^{-cmL}$, where
\ba c = -\int^{\infty}_0 {\frac{dx}{2\pi} \ln\left[\frac{x^2 + x\sqrt{x^2 
+1}}{ x^2 + 1 + x\sqrt{x^2 + 1}} \right]}. \ea
We conclude that the fidelity satisfies the scaling form $\ln F \sim -c L m$ 
which is in agreement with the prediction \ct{rams11} that $\ln F \sim -c L 
\de^{d \nu}$ where $\nu=d=1$ and $m=\de$ in the present case.

In case (ii), we can expand $\langle \psi (k_n, m) | \psi (k_n, -m)\rangle
= 1 ~-~ m^2 L^2/[2(2n+1)^2]$ to lowest order in $mL$. We then obtain
\be \ln F(m, -m) \simeq - \frac{m^2L^2}{2} ~\sum_{n=0}^\infty ~
\frac{1}{(2n+1)^2} = - \frac{\pi^2 m^2L^2}{16}. \ee
Hence we find the scaling relation $\ln F \sim \de^2 L^{2/\nu}$ in the 
non-thermodynamic limit; we can further conclude that $\chi_F \sim 
L^{2/\nu -d}$.
 
{ We now consider what happens if the fermions were interacting;
in one dimension, such a system is described by Tomonaga-Luttinger liquid 
theory \cite{mahan00,gogolin98,vondelft98,giamarchi04,giuliani05}. To be 
specific, let us consider the spin-1/2 $XXZ$ chain in a transverse magnetic 
field; the Hamiltonian is given by 
\ba H &=& \frac{1}{2} ~\sum_{n=-\infty}^\infty ~[~ \si_n^x \si_{n+1}^x ~+~ 
\si_n^y \si_{n+1}^y ~+~ J_z \si_n^z \si_{n+1}^z \non \\
& & ~~~~~~~~~~~~~- ~h_n \si_n^z ~]. \label{xxz} \ea
We first set $h_n =0$.
Upon using the Jordan-Wigner transformation which takes us from spin-1/2 to 
spinless fermions in one dimension \ct{lieb61}, we find that the first two 
terms in Eq.~(\ref{xxz}), $(1/2) (\si_n^x \si_{n+1}^x + \si_n^y \si_{n+1}^y)$,
lead to a tight-binding Hamiltonian of the form $- \sum_n (c_n^\dag c_{n+1} + 
c_{n+1}^\dag c_n)$; Fourier transforming to $k$-space, and linearizing around 
the two Fermi points lying at $k = \pm \pi/2$ gives the first term in
Eq. (\ref{ham1}). The third term in Eq. (\ref{xxz}), $\si_n^z \si_{n+1}^z$,
leads to a four-fermion interaction of the form $c_n^\dag c_n c_{n+1}^\dag 
c_{n+1}$. Eq. (\ref{xxz}) describes a system of massless interacting fermions 
if $J_z$ lies in the range $-1 \le J_z < 1$. (The cases $J_z =-1$ and $1$
describe an isotropic ferromagnet and isotropic antiferromagnet respectively).
The system is characterized by a Luttinger parameter $K$ given by
\be K ~=~ \frac{\pi}{2 \cos^{-1} (-J_z)}, \ee
so that $K$ goes from $\infty$ to $1/2$ as $J_z$ goes from $-1$ to 1. 
If $J_z =0$, there are no interactions between the fermions and we obtain 
$K=1$. 

We now introduce an alternating magnetic field of the form $h_n = m 
(-1)^n$; this introduces a coupling between the modes at the Fermi points 
$k = \pm \pi/2$ and therefore gives rise to the mass term in Eq. (\ref{ham1})
if $J_z = 0$ ($K=1$). The most efficient way of studying the low-energy,
long wavelength modes (i.e., the modes near the Fermi points)
of a one-dimensional system of
interacting fermions ($K \ne 1$) is to use the technique of bosonization 
\cite{mahan00,gogolin98,vondelft98,giamarchi04,giuliani05}.
Bosonization uses a quantum field theory which is defined in terms of 
a scalar field $\phi$. In this description, the action is given by
\be S ~=~ \frac{1}{2K} \int \int dt dx ~\left[~ \left( \frac{\partial \phi}{
\partial t} \right)^2 ~-~ \left( \frac{\partial \phi}{\partial x} \right)^2 
\right] \ee
for $m=0$ (we have again set the velocity equal to 1); the effect of 
interactions appears through the Luttinger parameter $K$. The contribution 
of the mass term to the action takes the form
\be S_m ~\sim~ \int \int dt dx ~m ~\cos (2 \sqrt{\pi} \phi). \ee
The operator $\cos (2 \sqrt{\pi} \phi)$ is known to have
mass dimension $K$; let us denote the coefficient of this operator in the
action by $\lambda$. It turns out that $\lambda$ effectively becomes
dependent on the length scale $L$, and $\lambda (L)$ satisfies the renormalization group (RG) equation
$d\lambda /d \ln L = (2-K) \lambda$. Given the initial value of $\lambda (a)
= m$ at some microscopic length scale $a$ (such as the lattice spacing),
and assuming that $m \ll 1$,
the RG equation implies that $\lambda (L)$ grows and becomes of order 1 at a
length scale given by $\xi$, where $\xi/a \sim 1/m^{1/(2-K)}$. Hence
the mass gap of the theory is given by $1/\xi \sim m^{1/(2-K)}$, 
leading to a low-energy dispersion given by $\omega_k = \sqrt{k^2 + m^{2/(2 
- K)}}$. One can then argue qualitatively that our above arguments about
fidelity would remain valid, but with a renormalized mass term given by 
$m^{1/(2-K)}$.  For case (i) where $L m^{1/(2-K)} \gg 1$, we would eventually 
find that the fidelity scales as $\ln F(m,-m) \sim -c^{'}L m^{1/(2 - K)}$, 
where $c'$ is a prefactor which differs from $c$ due to the presence of the 
interactions. Noting that the correlation length exponent in the presence of 
the mass perturbation is $\nu = 1/(2-K)$, we find that the scaling in 
Eq.~(\ref{fidelity_generic}) should also hold good, with $d=1$ and $\de = m$. 
The above analysis is only valid for $K < 2$; for $K > 2$, the mass 
perturbation is irrelevant in the sense of the renormalization group, 
and the fidelity has to be calculated in some other way which we will 
not pursue here.

We will not examine here the effects of a perturbation which takes the system 
across a Kosterlitz-Thouless transition; this occurs if $h_n = 0$ and $J_z$ 
crosses 1. The fidelity across this transition is considerably harder to 
analyze, and we refer to the work done by different groups 
\ct{yang07,fjaerstad08,chen08,sirker10,wang11}.}

\section{Kitaev Model}
 
In this section, we will exploit the solvability of the Kitaev model on the 
honeycomb lattice \ct{kitaev06} to calculate the fidelity of the model as a 
function of various parameters in the thermodynamic limit. We note that the 
fidelity per site studied for the same model has been able to detect quantum
phase transitions \ct{zhao09}, and the fidelity susceptibility has been 
calculated previously in the limit of small system sizes \ct{yang08,lin09}.

\subsection{Model, phase diagram and fidelity}

The Hamiltonian of the Kitaev model is given by 
\be H = \sum_{j+l=even} ~(J_1 \si_{j,l}^x \si_{j+1,l}^x + J_2 \si_{j-1,l}^y 
\si_{j,l}^y + J_3 \si_{j,l}^z \si_{j,l+1}^z), \label{ham_kit} \ee 
where $j$ and $l$ respectively denote the column and row indices of a 
honeycomb lattice (see Fig.~1) \ct{kitaev06,hikichi10,chen07,sengupta08}. We 
will 
assume that the couplings $J_i$, for $i=1,2,3$, are all positive; if some of 
them are negative, they can be made positive by appropriate $\pi$ rotations 
about the $x$, $y$ or $z$ spin axis. For the moment, our study will be 
restricted to the case $J_1=J_2$, although we will comment on the case 
with $J_1 \neq J_2$ in Sec. III E.

\begin{figure}[ht]
\begin{center}
\includegraphics[width=4.9cm]{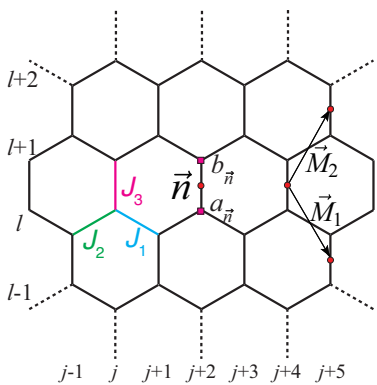}
\end{center}
\caption{(Color online) Schematic representation of the Kitaev model on a 
honeycomb lattice showing the bonds with couplings $J_1$, $J_2$ and $J_3$.
$\vec{M}_1$ and $\vec{M}_2$ are spanning vectors of the lattice. 
Sites `$a_{\vec{n}}$' and `$b_{\vec{n}}$' represent the two inequivalent sites
which make up a unit cell. (After reference [\onlinecite{hikichi10}])}
\label{Fig:hexagonal} \end{figure}

\begin{figure}[ht]
\begin{center}
\includegraphics[width=5.9cm]{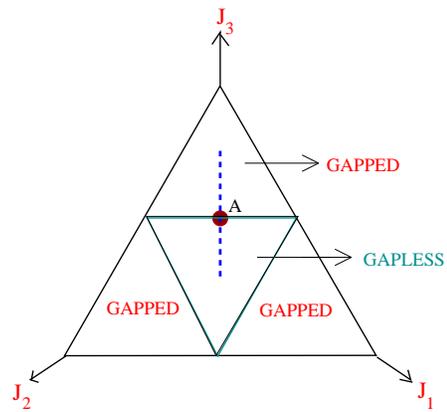}
\end{center}
\caption{(Color online) Phase diagram of the Kitaev model; all the points in 
the triangle satisfy $J_1 + J_2 + J_3 = 1$. The gapless phase is the region 
in which the couplings satisfy the triangle inequalities given by $J_1 \le 
J_2 + J_3$, $J_2 \le J_3 + J_1$ and $J_3 \le J_1 + J_2$, i.e., the points 
inside the inner equilateral triangle. Along the dashed vertical line $J_3$ 
is varied holding $J_1=J_2$, and the anisotropic quantum critical point (A) at
$J_3 = J_{3,c}=J_1+J_2 =2J_1$ is indicated. Our focus is to calculate the
fidelity between ground states lying on this vertical line.} \label{Fig:PD} 
\end{figure}

We define the Jordan-Wigner transformation as
\ba a_{j,l} &=& \left( \prod_{i = - \infty}^{j-1} \si_{i,l}^z \right) 
\si_{j,l}^y~~~\text{for even} ~~j+l, \non \\
a'_{j,l} &=& \left( \prod_{i = - \infty}^{j-1} \si_{i,l}^z \right) 
\si_{j,l}^x~~~\text{for even} ~~j+l, \non \\
b_{j,l} &=& \left( \prod_{i = - \infty}^{j-1} \si_{i,l}^z \right) 
\si_{j,l}^x~~~\text{for odd} ~~j+l, \non \\
b'_{j,l} &=& \left( \prod_{i = - \infty}^{j-1} \si_{i,l}^z \right) 
\si_{j,l}^y~~~\text{for odd} ~~j+l, \label{JWTransf} \ea
where $a_{j,l}$, $a'_{j,l}$, $b_{j,l}$ and $b'_{j,l}$ are all Majorana 
fermions, i.e., they are Hermitian, their square is equal to 1, and they
anticommute with each other. 
Instead of using the indices $(j,l)$ to specify the sites, we can use the 
two-dimensional vectors $\vec{n} = \sqrt{3} \hat{i} n_1 + (\frac{\sqrt{3}}{2} 
\hat{i} + \frac{3}{2} \hat{j}) n_2$ which denote the midpoints of the vertical
bonds of the honeycomb lattice; {here $\hat i$ denotes the unit 
vector along the horizontal (labeled by $j, ~j+1$, etc) and 
similarly $\hat j$ is the unit vector along the vertical direction}. 
Here $n_1$ and $n_2$ run over all integers so 
that the vectors $\vec n$ form a triangular lattice. The Majorana fermions 
$a_{\vec n}$ ($a'_{\vec n}$) and $b_{\vec n}$ ($b'_{\vec n}$) are located at 
the bottom and top lattice sites respectively of the bond labeled by $\vec n$.
The vectors $\vec{M_1} = \frac{\sqrt{3}}{2} \hat{i} - \frac{3}{2} \hat{j}$ and
$\vec{M_2} = \frac{\sqrt{3}}{2} \hat{i} + \frac{3}{2} \hat{j}$ shown in Fig.~1
are the spanning vectors of the lattice. (We have set the nearest neighbor 
lattice spacing to unity).

The Fourier transforms of the Majorana fermions are given by
\be a_{\vec n} ~=~ \sqrt{\frac{4}{L}} ~\sum_{\vec k} ~[~ a_{\vec k} ~e^{i
{\vec k} \cdot {\vec n}} ~+~ a_{\vec k}^\dg ~ e^{-i{\vec k} \cdot {\vec n}}~],
\label{ft} \ee
satisfying $\{ a_{\vec k}, a^\dg_{\vec k'} \} = \de_{{\vec k},
{\vec k'}}$, and similarly for $a'_{\vec n}$, $b_{\vec n}$ and $b'_{\vec n}$.
In Eq. (\ref{ft}), $L$ is the number of sites (hence the number of unit cells 
is $L/2$), and the sum over $\vec k$ extends over {\it half} the Brillouin 
zone of the hexagonal lattice because of the Majorana nature of the fermions 
\ct{chen07,sengupta08}. The full Brillouin zone is given by a rhombus with 
vertices lying at $(k_x,k_y)=(\pm2 \pi /\sqrt{3},0)$ and $(0,\pm 2 \pi /3)$;
half the Brillouin zone is given by an equilateral triangle with vertices
at $(k_x,k_y)=(2 \pi /\sqrt{3},0)$ and $(0,\pm 2 \pi /3)$. 

In terms of the Majorana fermions, the Hamiltonian in Eq. (\ref{ham_kit})
takes the form
\be H^{\prime} = i \sum_{\vec n} \left( J_1 b_{\vec n}a_{\vec{n}- \vec{M_1}}
+ J_2 b_{\vec n}a_{\vec{n}+ \vec{M_2}} + J_3 D_{\vec n} b_{\vec n} a_{\vec n}
\right), \label{H2} \ee
where $D_{\vec n} = i ~b'_{\vec n} a'_{\vec n}$. We note that the 
operators $D_{\vec n}$ have eigenvalues $\pm 1$, and commute with each other 
and with $H^{\prime}$; hence all the eigenstates of $H^{\prime}$ can be 
labeled by specific values of $D_{\vec n}$. (We observe that 
the Hamiltonian $H^{\prime}$ gives dynamics to the fermions $a_{\vec n}$
and $b_{\vec n}$, but the fermions $a'_{\vec n}$ and $b'_{\vec n}$ have no 
dynamics since $i b'_{\vec n} a'_{\vec n}$ is fixed). The ground state can be 
shown to correspond to $D_{\vec n} = 1$ for all ${\vec n}$. \ct{kitaev06} For 
$D_{\vec n} = 1$, the Hamiltonian can be diagonalized into the form 
\be H^{\prime} = \sum_{\vec k}~ \left( \begin{array}{cc}
a^\dg_{\vec k} & b^\dg_{\vec k} \end{array} \right) ~H_{\vec k} ~
\left( \begin{array}{c}
a_{\vec k} \\
b_{\vec k} \end{array} \right), \ee
where $H_{\vec k}$ can be written in terms of Pauli matrices as
\ba H_{\vec k} &=& \al_{\vec k} ~\si^1 ~+~\beta_{\vec k} ~\si^2, \non \\
{\rm where}~~ \al_{\vec k} &=& 2 [ J_1 \sin(\vec{k} \cdot \vec{M_1}) - J_2 
\sin(\vec{k} \cdot \vec{M_2})], \non \\
{\rm and} ~~\beta_{\vec k} &=& 2 [ J_3 + J_1 \cos(\vec{k} \cdot \vec{M_1}) 
+ J_2 \cos(\vec{k} \cdot \vec{M_2})]. \non \\
& & \label{hamilreduced} \ea
The energy spectrum of $H^{\prime}$ consists of two bands with energies 
given by
\be E^{\pm}_{\vec k} ~=~ \pm ~ \sqrt{\al^2_{\vec k} ~+~ \beta^2_{\vec k}}.
\label{spectrum} \ee
The energy gap $E^+_{\vec k}-E^-_{\vec k}$ vanishes for specific values of 
$\vec k$ when $|J_1-J_2| \le J_3 \le J_1+J_2$ giving rise to a gapless phase 
of the model. The gapless and gapped phases of the model are shown in 
Fig.~\ref{Fig:PD} in terms of points in an equilateral triangle which satisfy 
$J_1 + J_2 + J_3 = 1$ and all $J_i > 0$, with the value of $J_i$ being given 
by the distance from the opposite side of the triangle as indicated by the 
arrows. In the limit $J_3=0$, the Hamiltonian (\ref{ham_kit}) reduces to 
a one-dimensional version of the Kitaev model \ct{feng07} which in turn can be
mapped to the transverse Ising chain following a duality transformation 
\ct{capel77}.

In particular, let us consider the critical line $J_3 = J_1 + J_2$ which 
separates one of the gapped phases from the gapless phase in Fig.~\ref{Fig:PD}.
On this line, the energy vanishes at the three corners of half the Brillouin 
zone given by ${\vec k} = (2\pi/\sqrt{3},0)$ and $(0, \pm 2\pi /3)$; these
three points are actually equivalent to each other because they are related by
shifts by the lattice vectors $G_{\pm} = (2\pi/\sqrt{3},\pm 2\pi /
3)$. If $(dk_x,dk_y)$ denotes a small deviation from any one of these three 
points, we find that the energy is highly anisotropic with respect to this
deviation. Namely, for $J_3 = J_1 + J_2$, the quantities $\al_{\vec k}$ and 
$\beta_{\vec k}$ appearing in Eqs.~(\ref{hamilreduced}-\ref{spectrum}) are 
given by 
\ba \al_{\vec k} &=& \sqrt{3} (J_2 - J_1) dk_x + 3 (J_1 + J_2) dk_y, \non \\
\beta_{\vec k} &=& J_1 \left( \frac{\sqrt{3}}{2} dk_x - \frac{3}{2} dk_y 
\right)^2 + J_2 \left( \frac{\sqrt{3}}{2} dk_x + \frac{3}{2} dk_y \right)^2 , 
\non \\
& & \label{disp} \ea
respectively, to lowest order in $dk_x$ and $dk_y$. We see that $\al_{\vec k}$
varies linearly in one particular direction in the plane of $(dk_x,dk_y)$,
while $\beta_{\vec k}$ varies quadratically in any direction. We thus have 
an AQCP \ct{hikichi10}. For simplicity, we will mainly restrict our attention 
below to the case where $J_1 = J_2$ is held fixed and $J_3$ is varied along 
the dashed vertical line shown in Fig.~\ref{Fig:PD}. Then the point 
$J_3=J_{3,c} = 2J_1$ marked by $A$ is an AQCP, with the energy gap vanishing 
near the three points as $E_{\vec k} \sim (dk_x)^2$ and $dk_y$ for deviations 
along the $k_x$ and $k_y$ directions respectively. {For $J_1 =
J_2$, the dispersion is linear along the vertical direction $\hat j$ and
quadratic along the horizontal direction $\hat i$ 
(see Fig.~\ref{Fig:hexagonal}).
This implies, for the analysis given in Sec. III B, that the correlation length
exponent $\nu_{\perp} = 1$ and $L_{\perp}$ is the length of the system in the 
$\hat j$ direction, while the exponent $\nu_{||} = 1/2$ and $L_{||}$ is 
the length of the system in the $\hat i$ direction. For a more general
AQCP given by $J_3 = J_1 + J_2$ but $J_1 \ne J_2$, Eq. (\ref{disp})
implies that the dispersion is linear along a direction given by ${\hat e}_1 =
\sqrt{3} (J_2 - J_1) {\hat i} + 3 (J_1 + J_2) {\hat j}$ and quadratic in a 
direction ${\hat e}_2$ which is perpendicular to ${\hat e}_1$. Hence 
$\nu_{\perp} = 1$ and $L_{\perp}$ is the length of the system in the 
${\hat e}_1$ direction, while $\nu_{||} = 1/2$ and $L_{||}$ is the length 
of the system in the ${\hat e}_2$ direction.}

We will now show that the ground state of the model in Eq. (\ref{H2}) can 
be written as a product over all $\vec k$ lying in half the Brillouin zone. 
Firstly, the unprimed Majorana fermions $a_{\vec k}$ and $b_{\vec k}$ must 
be chosen to have the lower eigenvalue $E^-_{\vec k}$ of $H_{\vec k}$; the 
corresponding normalized state is given by 
\ba |S_{\vec k} \rangle &=& (1/\sqrt{2}) ~(~ a^\dg_{\vec k} ~-~ 
e^{i\theta_{\vec k}} ~b^\dg_{\vec k}) ~|\Phi \rangle, \non \\
{\rm where}~~ e^{i\theta_{\vec k}} &=& \frac{\al_{\vec k} ~+~ i 
\beta_{\vec k}}{\sqrt{\al^2_{\vec k} ~+~ \beta^2_{\vec k}}}, \label{sk} \ea
and $|\Phi \rangle$ is the vacuum state annihilated by $a_{\vec k}$, 
$a'_{\vec k}$, $b_{\vec k}$ and $b'_{\vec k}$. Secondly, the condition 
$D_{\vec n} 
= i b'_{\vec n} a'_{\vec n} = 1$ for all $\vec n$ implies that if we define 
the Dirac fermion operators $c_{\vec n} = (1/2) (a'_{\vec n} - i b'_{\vec n})$,
the ground state must be an eigenstate of $c^\dg_{\vec n} c_{\vec n}$ with 
eigenvalue 1 for all $\vec n$. Hence the state must be annihilated by
$c^\dg_{\vec n}$ for all $\vec n$; taking the Fourier transform of this 
means that the state must be annihilated by both $c^\dg_{\vec k} =(1/2)
(a'^\dg_{\vec k} + i b'^\dg_{\vec k})$ and $c_{\vec k} = (1/2)
(a'_{\vec k} + i b'_{\vec k})$ for all $\vec k$. Hence the normalized state 
is given by
\be |T_{\vec k} \rangle ~=~ (1/\sqrt{2})~ (~a'^\dg_{\vec k} ~+~ i ~
b'^\dg_{\vec k}) ~|\Phi \rangle \label{tk} \ee
for each $\vec k$. The complete ground state is therefore given by the product 
\be | \Psi \rangle ~=~ \prod_{\vec k} ~\left[ \frac{1}{2} ~(a^\dg_{\vec k}
- e^{i\theta_{\vec k}} ~b^\dg_{\vec k})~(~a'^\dg_{\vec k} + i ~
b'^\dg_{\vec k}) \right] ~|\Phi \rangle. \label{prod1} \ee
Using Eq. (\ref{prod1}), we can write the ground state fidelity in the 
form \ct{yang08}
\ba F^2 &=& \prod_k| \langle \Psi^+|\Psi^- \rangle|^2 = \prod_k 
\frac{1}{2} \left(1 + \frac{\al_{\vec k}^+ \al^-_{\vec k} + \beta_{\vec k}^+
\beta_{\vec k}^-}{E_{\vec k}^+ E_{\vec k}^-} \right) \non \\ 
&=& \prod_k {\cos^2\left(\frac{\theta_{\vec k}^+ - \theta_{\vec k}^-}{2} 
\right)}, \label{fidelity_prod} \ea
where
\be \cos{\theta_{\vec k}^{\pm}} ~=~ \frac{\al_{\vec k}^{\pm}}{
E_{\vec k} ^{\pm}} ~~~{\rm and} ~~~ \sin{\theta_{\vec k}^{\pm}} ~=~ 
\frac{\beta_{\vec k}^{\pm}}{E_{\vec k}^{\pm}}, \ee
with the $\pm$ in the superscripts denoting the corresponding values with $J_3 
\pm \de$. One finds 
\ba \ln F \simeq {\de^2 L^2} \int^{\pi-\pi/L}_{\pi/L} \int^{\pi-\pi/L}_{\pi/L}
dk_x dk_y ~\frac{\alpha_{\vec k}^2}{\alpha_{\vec k}^2 + \beta_{\vec k}^2}. 
\label{fidelity_full} \ea
Analyzing for small $\de$ close to the AQCP, we find
\ba \ln F \approx -\frac{9\de^2 L^2}{2 \pi^2} \int^{\infty}_{\pi/L} 
\int^{\infty}_{\pi /L} \frac{k_y^2 dk_x dk_y}{R_+ R_-}, 
\label{fidelity_analytical} \ea
where $R_{\pm} = 9k_y^2 + \left(\frac{3}{4}k_x^2 - \la \pm \de \right)^2$, and
we have only included contributions coming from the low energy modes close to 
the critical modes and extended the limit of integrations to $\infty$.

In subsequent sections, we will investigate the fidelity between the two 
ground states of the model with interaction terms $J_3 = J_{3,c} - \la + \de$
and $J_3 = J_{3,c} - \la - \de$, respectively, with $J_1 = J_2$, i.e., along 
the vertical line in Fig.~\ref{Fig:PD}; here $\la$ 
and $\de$ 
determine the location in the phase diagram. 

We will use the simplified equation (\ref{fidelity_analytical}) to derive the
scaling of fidelity analytically. On the other hand, for the purpose of 
numerical analysis of Eq. (\ref{fidelity_full}), we will parametrize the 
momenta $k_x$ and $k_y$ in terms of two independent variables $v_1$ 
and $v_2$ for $0 \le v_1,v_2 \le 1$, given by 
\be k_x = \frac{2\pi}{\sqrt 3}~(v_1 +v_2 -1) ~~~{\rm and}~~~ k_y = 
\frac{2\pi}{3} ~(v_1 -v_2), \label{v1v2} \ee
which ensures that all the points in the rhombus are covered uniformly. Once 
again, we need to avoid the corners of the Brillouin zone (i.e, the values 0 
and 1 for $v_1$ and $v_2$), otherwise the fidelity will turn out to be zero. 
We will let $v_1$ and $v_2$ go from $1/(2L)$ to $1-1/(2L)$ in steps of $1/L$, 
where $L$ is a large integer. Finally, we must take $v_1 + v_2 \ge 1$ so as 
to restrict the integral to half the Brillouin zone.

\subsection{General scaling of fidelity near an AQCP}

We will now proceed to derive a scaling form for the fidelity in the 
thermodynamic limit near a $d$-dimensional generic AQCP in the same spirit as 
in Ref. \onlinecite{rams11}. The 
corresponding scalings in the limit of small system size is given in Ref. 
\onlinecite{mukherjee11}. We consider a situation in which the correlation 
length exponent and system size are given by $\nu = \nu_{||}$ and $L = 
L_{||}$, respectively, along $m$ spatial dimensions, and $\nu = \nu_{\perp}$
and $L = L_{\perp}$, respectively, along the remaining $d - m$ dimensions. We 
encounter such a case with $d=2, m=1$, $\nu_{||} = 1/2$ and $\nu_{\perp} = 1$
in the two-dimensional Kitaev model (point (A) in the phase diagram) and also 
near a semi-Dirac band crossing point \ct{banerjee09}. We consider the scaling
parameter \ct{zhou083}
\ba & & S(\la + \de, \la - \de) \non \\
&=& -\lim_{N \to \infty} \frac{\ln{\left|\langle \psi_0 (\la + \de
)|\psi_0( \la - \de) \rangle\right|}}{N} \non \\ 
&=& -\lim_{N\to \infty} \frac{\ln{F(\la - \de, \la + \de)}}{N}, \ea
where $N = L_{||}^m L_{\perp}^{d - m}$ is the system size, $\la$ is the 
distance from the AQCP, and $\la$, $\de$ are assumed to be 
positive. We propose the scaling ansatz
\ba && S(\la + \de, \la - \de) ~= \non \\
& &  L_{||}^{-m}L_{\perp}^{-(d - m)} ~f ( (\la + \de) L_{||}^{1/\nu_{||}}, 
(\la + \de) L_{\perp}^{1/\nu_{\perp}}, \non \\
& & (\la - \de) L_{||}^{1/\nu_{||}}, (\la -\de) L_{\perp}^{1/\nu_{\perp}}), \ea
where $f$ is a scaling function that is symmetric with respect to the 
operation $\de \to -\de$. Rescaling $L_{||}$($L_{\perp}$) to 
$b_{||}$ ($b_{\perp}$) and choosing $b_{||}$, $b_{\perp}$ such that $(\la 
+ \de) b_{||}^{1/\nu_{||}} = (\la + \de) 
b_{\perp}^{1/\nu_{\perp}} = 1$, we get
\ba && S(\la + \de, \la - \de) \non \\
&=& (\la + \de)^{\nu_{||}m + \nu_{\perp}(d - m)}f\left(1, 
\frac{\la - \de}{\la + \de} \right). \label{ansatz} \ea
Taking the limit $\de/\la \to 0$, and expanding $f\left(1, 
\frac{1 - \de/\la}{1 + \de/\la} \right) = g (\de/\la)$ around $\de/\la = 0$, 
we arrive at the scaling form
\ba &&\ln{F}(\la + \de, \la - \de) \non \\ 
&\sim& -\de^2 L_{||}^m L_{\perp}^{d - m}\la^{\nu_{||}m + 
\nu_{\perp}(d - m) - 2}, \label{aqcp_la} \ea
where we have taken $g(0) = g'(x)|_{x = 0}$. Now let us focus on the case 
$\la = 0$, i.e., we are studying the fidelity between two states at $\de$ and 
$-\de$, respectively, on either side of the AQCP. In the limit $\la = 0$, 
Eq.~(\ref{ansatz}) shows that
\ba \ln{F}(\de,-\de) \sim -L_{||}^m L_{\perp}^{d - m} 
\de^{\nu_{||}m + \nu_{\perp}(d - m)}. \label{general_scaling_fidelity} \ea
We note that the above scaling forms are valid only as long as the 
corresponding exponent of $\la$ (see Eq. (\ref{aqcp_la})) or 
$\de$ (see Eq. (\ref{general_scaling_fidelity})) does not exceed 2.
Otherwise the low-energy singularities associated with the critical point 
become subleading to the quadratic scaling form of perturbation theory, 
and $|\ln{F}|$ starts varying as $\la^2$ (or as $\de^2$ if $\la = 0$) instead,
irrespective of the critical exponents \ct{gritsev09}. Both 
Eqs.~(\ref{aqcp_la}) and (\ref{general_scaling_fidelity}) reduce to the 
scaling presented in Ref. \onlinecite{rams11} for $\nu_{||}=\nu_{\perp}= \nu$.

Now we will consider Eq. (\ref{general_scaling_fidelity}) in the 
non-thermodynamic limit ($\de \ll L_{\perp}^{-1/\nu_{\perp}}$) and choose 
$L_{\perp}^{-1/\nu_{\perp}} > L_{||}^{-1/\nu_{||}}$. In this limit, a 
cross-over from a dependence on $\de $ to 
a dependence on the $L_{\perp}$ takes place in 
the scaling in Eq.~(\ref{general_scaling_fidelity})
which then takes the form \ct{mukherjee11}
\ba \ln{F}(\de,-\de) &\approx& -\de^2 L_{||}^m L_{\perp}^{d - m} \chi_F \non \\
&\sim& -\de^2 L_{||}^mL_{\perp}^{\frac{2}{\nu_{\perp}} - 
\frac{\nu_{||}}{\nu_{\perp}}m}, \label{general_scaling_fidelity_nonthermo} \ea
where the $\de^2$ in Eq.~(\ref{general_scaling_fidelity_nonthermo}) arises due 
to perturbation theory. In contrary, when $\de \ll L_{||}^{-1/\nu_{||}}$ and 
$L_{||}^{-1/\nu_{||}} > L_{\perp}^{-1/\nu_{\perp}}$, we get 
\ba \ln{F}(\de,-\de) &\sim& -\de^2 L_{\perp}^{(d - m)}L_{||}^{
\frac{2}{\nu_{||}} - \frac{\nu_{\perp}}{\nu_{||}}(d - m)}. \ea

We will now verify the above scaling for the AQCP (A) shown in 
Fig.~\ref{Fig:PD} and determine the fidelity between the two ground states 
at $J_3 = J_{3,c} + \de$ and $J_3 = J_{3,c} - \de $ with $J_1=J_2$;
the system lies in the gapless phase for $J_3 = J_{3,c} - \de$ and in the 
gapped phase for $J_3 = J_{3,c} + \de$. For all numerical studies presented
hereafter we have set $L_{||}=L_{\perp}=L$. 

\begin{figure}[htb]
\includegraphics[height=2.2in,width=3in, angle = 0]{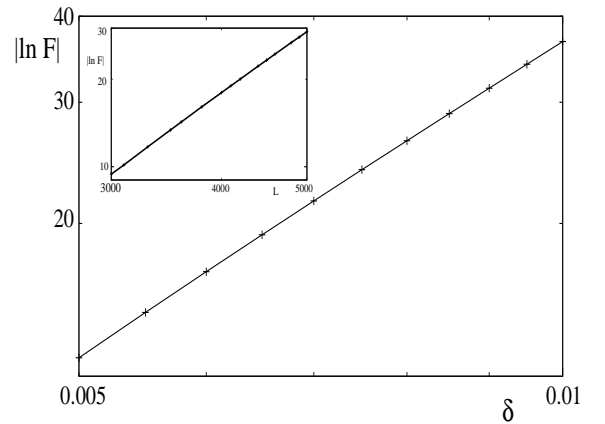}
\caption{Variation of $|\ln{F}|$ with $\de$ as obtained numerically at the 
AQCP in the thermodynamic limit for $J_1 = J_2 = 1$, $L = 1001$ and $\la 
= 0$. $|\ln{F}|$ varies as $\de^{3/2}$. Inset: Variation of $|\ln{F}|$ with
$L$ for $\de = 0.001$ and $\la = 0$. $|\ln{F}|$ varies as $L^2$.}
\label{fig_thermo_aqcp} \end{figure}

We use Eq. (\ref{fidelity_analytical}) to arrive at the scaling relations 
followed by the quantum fidelity; rescaling $k_x \to k_x^{'} = k_x/\sqrt{\de}$
and $k_y \to k_y^{'} = k_y/{\de}$, we get
\ba \ln{F} &\approx& -\frac{9\de^{3/2} L^2}{2\pi^2}\int_{\pi/L
\sqrt{\de}}^{\infty}\int_{\pi/L\de}^{\infty}\frac{{k_y^{'}}^2 
dk_x^{'}dk_y^{'}}{R_{+}^{'}R_{-}^{'}} \non \\
&\approx& -\frac{9\de^{3/2} L^2}{2\pi^2}\int_{0}^{\infty}\int_{0}^{\infty}
\frac{{k_y^{'}}^2 dk_x^{'}dk_y^{'}}{R_{+}^{'}R_{-}^{'}} \non \\
&\sim& -\de^{3/2} L^2 \label{scaling_del} \ea
in the limit $\de \gg L^{-1/\nu_{\perp}} = L^{-1}$, as expected from Eq. 
(\ref{general_scaling_fidelity}) (see Fig.~\ref{fig_thermo_aqcp} for numerical
verification). In the above Eq. (\ref{scaling_del}) we have taken 
$R_{\pm}^{'} = \left[9{k_y^{'}}^2 + \left(\frac{3}{4}{k_x{'}}^2 \pm 1 
\right)^2 \right]$. 

\begin{figure}[htb]
\includegraphics[height=2.2in,width=3in, angle = 0]{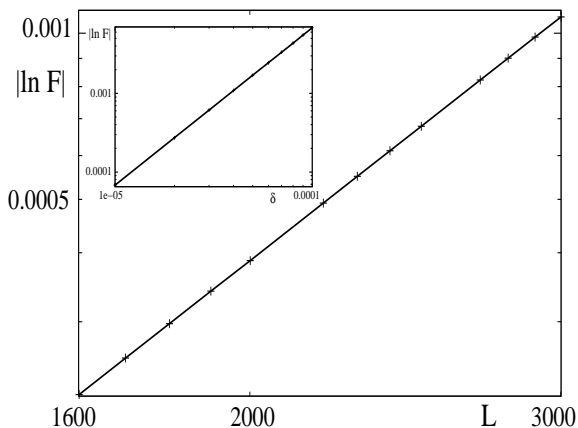}
\caption{Variation of $|\ln{F}|$ with $L$ as obtained numerically at the AQCP 
in the non-thermodynamic limit for $J_1 = J_2 = 1$, $\de = 0.00001$ and 
$\la = 0$. $|\ln{F}|$ varies as $L^{5/2}$. Inset: Variation of $|\ln{F}|$ 
with $\de$ as obtained numerically at the AQCP in the non-thermodynamic 
limit for $J_1 = J_2 = 1$, $L = 1001$ and $\la = 0$ . $|\ln{F}|$ varies 
as $\de^2$.} \label{fig_nonthermo_aqcp} \end{figure}

In the non-thermodynamic limit of $\de \ll L^{-1}$, on the other hand, we 
can use the transformation $k_x = q\sqrt{k_y}$ to arrive at the scaling 
\ba \ln F \approx -\frac{9\de^2 L^2}{2\pi^2}\int_{\pi/L}^{\infty}
\int_{\pi/L\sqrt{k_y}}^{\infty}\frac{k_y^2 \sqrt{k_y}dk_y dq }{k_y^4 P_{+}
P_{-}}, \label{fidelity_P} \ea
where $P_{\pm} = 9 + \left(3q^2/4 \pm \de/k_y \right)^2$. Now, our scaling 
transformation suggests $q \sim 1/\sqrt{L} \ll 1$ and also the limit 
$\de \ll L^{-1}$ implies $\de/k_y \ll 1$.
The above analysis shows that for small values of $q$ and $k_y$, which give 
the dominant contributions to the integral in Eq. (\ref{fidelity_P}), 
$P_{\pm}$ are of the order of unity. Therefore we get
\ba \ln F &\sim& - \de^2 L^2\int^{\infty}_{\pi/L} \frac{d k_y}{k_y^{3/2}}
\int^{\infty}_{\pi/L\sqrt{k_y}}\frac{d q}{P_{+}P_{-}} \non \\
&\sim& - \de^2 L^2\int^{\infty}_{\pi/L} \frac{d k_y}{k_y^{3/2}} \non \\
&\sim& - \de^2 L^2 \chi_F (J_3 = J_{3,c}) \sim -\de^2 L^{5/2},
\label{kitaev_F} \ea
which is in complete agreement with our prediction in 
Eq.~(\ref{general_scaling_fidelity_nonthermo}), as shown in 
Fig.~\ref{fig_nonthermo_aqcp}. We note that earlier studies of the fidelity 
susceptibility in the thermodynamic limit in the two-dimensional Kitaev model 
have pointed to the same scaling form as in Eq.~(\ref{kitaev_F}) 
\ct{yang08,lin09}.

We reiterate that the study of the scaling of fidelity in the thermodynamic 
limit is closely related to that of fidelity per site 
\ct{zhou082,zhou083,zhao09}. The quantum phase transition at $J_3 =J_{3,c}$ 
is associated with a singularity in the double derivative
of the scaling function $S = -\lim_{N \to \infty} \ln F/N$ given by
\ba \frac{\partial^2 \ln S}{\partial J_3^2} = C \ln |J_3 - J_{3,c}| + 
\text{constant}, \ea
where $C$ is a negative constant and hence one observes a dip close to the 
QCP, as shown in Ref. \onlinecite{zhao09}.

\begin{figure}[htb]
\includegraphics[height=2.2in,width=3.3in, angle = 0]{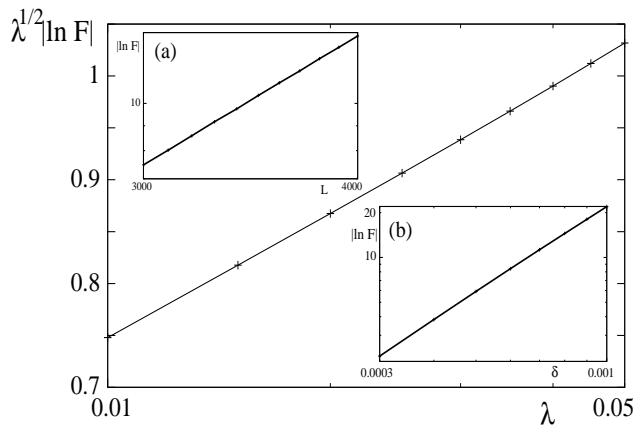}
\caption{Variation of $\la^{1/2}|\ln{F}|$ with $\la$ as obtained 
numerically inside the gapless region in the limit $\de L \gg 1$, for $J_1 = 
J_2 = 1$, $\de = 0.001$ and $L = 3001$. $|\ln{F}|$ varies as 
$\la^{-1/2}\ln{\la}$. Inset: (a) Variation of $|\ln{F}|$ with $L$ as 
obtained numerically inside the gapless region for 
$J_1 = J_2 = 1$, $\de = 0.001$ and $\la = 0.01$. $|\ln{F}|$ varies as 
$L^2$. (b) Variation of $|\ln{F}|$ with $\de$ as obtained numerically 
inside the gapless region for $J_1 = J_2 = 1$, 
$L = 5001$ and $\la = 0.01$. $|\ln{F}|$ varies as $\de^2$.}
\label{fig_gapless_thermo} \end{figure}

\begin{figure}[htb]
\includegraphics[height=2.2in,width=3.3in, angle = 0]{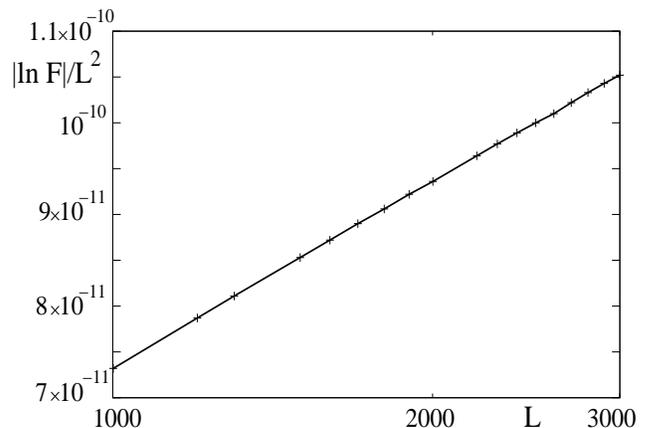}
\caption{Variation of $|\ln{F}|/L^2$ with $L$ as obtained numerically inside 
the gapless region in the limit $\de L \ll 1$, for $J_1 = J_2 = 1$, 
$\de = 0.00001$ and $\la = 0.005$. $|\ln{F}|$ varies as $L^2\ln{L}$.}
\label{fig_gapless_nonthermo_L} \end{figure} 

\subsection{Fidelity inside the gapless region:}

In this section, we consider the situation when both the states under 
consideration lie inside the gapless region 
of the phase diagram (Fig.~\ref{Fig:PD}) along the dashed vertical line 
with $\la= J_3 -J_{3,c}\gg \de >
0$ and $\la \gg L^{-1/\nu_{\perp}}=L^{-1}$. To calculate quantum
fidelity, we numerically integrate Eq.~(\ref{fidelity_full}) and arrive 
at the scaling relation
\ba \ln F \sim -\de^2 L^2 \la^{-1/2}\ln\la 
\label{fidelity_gapless_thermodynamic} \ea
in the limit $\de L^{1/\nu_{\perp}} = \de L \gg 1$. In 
Figs.~(\ref{fig_gapless_thermo}), we present the numerical results which 
clearly support the above scaling prediction.

Close to the AQCP, one can provide an analytical verification of 
(\ref{fidelity_gapless_thermodynamic}) using Eq. 
(\ref{fidelity_analytical}) with $R_{\pm}$ as defined before. Using the 
transformations $k^{'}_x = k_x/\sqrt{\la}$, $k_y^{'} = k_y/\la$, Eq. 
(\ref{fidelity_analytical}) can be rewritten as
\ba & & \ln F \non \\
&\approx& -\frac{9\de^2 L^2 \la^{-1/2}}{\pi^2} \int^{\infty}_{\pi/L\la} 
\int^{\infty}_{\pi/L\sqrt{\la}} \frac{{k^{'}_y}^2 dk^{'}_x dk^{'}_y}{
\left[9{k^{'}_y}^2 + \left(\frac{3}{4}{k^{'}_x}^2 -1\right)^2\right]^2} \non \\
&\sim& \de^2 L^2 \la^{-1/2} g(L,\la), \label{fidelity_analytical2} \ea
in the limit $\la \gg \de$ when $\de$ appearing in the integrand can
be ignored. The function $g(L,\la)$ is found to scale as $g(L, \la) 
\sim \ln \la$ by numerical investigations of Eq.~(\ref{fidelity_full}). We 
interpret this logarithmic behavior in (\ref{fidelity_gapless_thermodynamic}) 
as a signature of the system being in the gapless region.
The scaling $|\ln F| \sim \la^{-1/2}\ln\la$ can be understood noting 
that $\la$ denotes the distance from the AQCP; the power-law scaling
$\la^{-1/2}$ follows from the generic scaling in Eq.~(\ref{aqcp_la}), while 
the gapless nature of the phase diagram is encoded in the additional 
logarithmic correction. 

On the other hand, in the limit $\de L \ll 1$, again with $\la \gg L^{-1}$, a 
similar analysis of Eq.~(\ref{fidelity_full}) leads to the scaling
\ba \ln F &\sim& -\de^2 L^2 \la^{-1/2}\ln{L} \ln{\la},\ea
as shown in Fig.~\ref{fig_gapless_nonthermo_L}; we therefore find an 
additional $\ln L$ correction in comparison to the scaling in
(\ref{fidelity_gapless_thermodynamic}).
Interestingly, it can be shown that there exists another cross-over at 
$\la \lesssim L^{-1/\nu_{\perp}} = L^{-1}$, when the system size dependence 
changes to $\ln F \sim L^{5/2}$. This is expected as the system approaches 
the vicinity of AQCP where the scaling Eq. 
(\ref{general_scaling_fidelity_nonthermo}) is applicable.

A few comments are necessary at this point. Our analysis points to  a 
cross-over from $\ln F \sim L^2$ for $\de L^{1/\nu_{\perp}}=\de L \gg 1$ to 
$\ln F \sim L^2 \ln L$ for $\de L \ll 1$ with $\la \gg L^{-1/\nu_{\perp}}$ in 
both the cases (see Eqs.~(43) and (45), above). This apparently suggests that even in the gapless phase, we 
see a cross-over around $\de L \sim 1$, which resembles a thermodynamic to 
non-thermodynamic cross-over in fidelity, as observed in Ref. 
\onlinecite{rams11}, 
though scaling with $\de$ remains the same in the present case. It also 
appears that the crossover occurs as $\de L^{1/\nu_{\perp}} \sim 1, $, which 
suggests that the AQCP may play the role of a dominant critical point in its 
vicinity in the gapless phase.
 

\subsection{Fidelity inside the gapped phase: }

For $\la < 0$, both the states are in the gapped phase along the vertical 
line of Fig.~\ref{Fig:PD}, and choosing $\la \gtrsim L^{-1}$ and $\la \gg 
\de$ (i.e., non-thermodynamic limit), one finds numerically 
\ba \ln F \sim -\de^2 L^d \la^{\nu_{||}m + \nu_{\perp}(d - m) - 2} 
\sim \de^2 L^2 \la^{-1/2}, \ea
which matches exactly with non-thermodynamic result for fidelity 
susceptibility\ct{lin09}. As discussed above, we encounter a cross-over to 
$|\ln F| \sim L^{5/2}$ for $\la \lesssim L^{-1}$.

\subsection{Observations for $J_1 \neq J_2$}

\begin{figure}[htb]
\includegraphics[height=2.2in,width=3.3in, angle = 0]{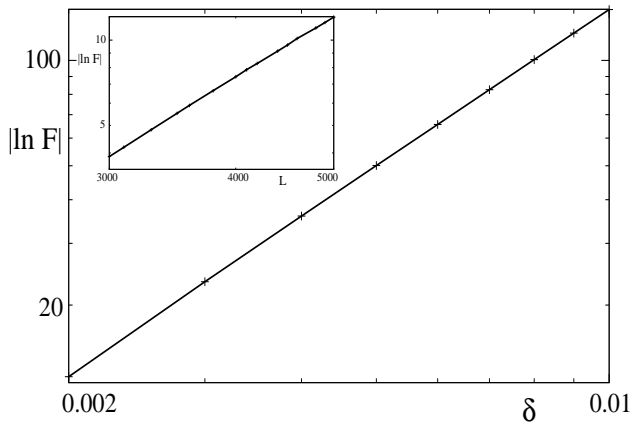}
\caption{Variation of $|\ln{F}|$ with $\de$ in the thermodynamic limit, as
obtained numerically for $J_3 = J_1 + J_2$, $J_1 = 3J_2 = 3$, $L = 3001$ and 
$\la = 0$. $|\ln{F}|$ scales as $\de^{3/2}$. Inset: Variation of 
$|\ln{F}|$ with $L$ in the thermodynamic limit, as obtained numerically for 
$J_3 = J_1 + J_2$, $J_1 = 3J_2 = 3$, $\la = 0$ and $\de = 0.001$. 
$|\ln{F}|$ shows a quadratic scaling with $L$ in this limit.} 
\label{fig_thermo_iso} \end{figure}

\begin{figure}[htb]
\includegraphics[height=2.2in,width=3.3in, angle = 0]{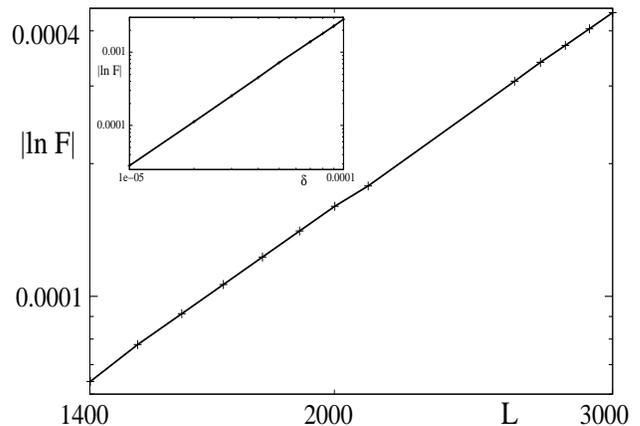}
\caption{Variation of $|\ln{F}|$ with $L$ in the non-thermodynamic limit, as
obtained numerically for $J_3 = J_1 + J_2$, $J_1 = 3J_2 = 3$, $\la = 0$ 
and $\de = 0.00001$. $|\ln{F}|$ varies as $L^{5/2}$ in this regime. Inset: 
Variation of $|\ln{F}|$ with $\de$ in the non-thermodynamic limit, as 
obtained numerically for $J_3 = J_1 + J_2$, $J_1 = 3J_2$, $\la = 0$ and 
$L = 1001$. $|\ln{F}|$ scales quadratically with $\de$.} 
\label{fig_nonthermo_iso} \end{figure}

Our attention so far has been concentrated on the case $J_1=J_2$. However, 
all the points on the critical line $J_3 = J_1 + J_2$ correspond to an AQCP, 
regardless of whether $J_1 = J_2$ or not. In Figs.~\ref{fig_thermo_iso} - 
\ref{fig_nonthermo_iso}, we have presented our numerical results with 
$J_1=3J_2$ which clearly shows the effect of the AQCP.

\section{Calculating fidelity in the Kitaev model using rotation of spins}

In this section, we will compute the overlap between two ground state wave 
functions of the Kitaev model with each spin rotated about some axis by an 
angle $\eta$ and $\eta+d\eta$, respectively; we note that a similar method has
been used to calculate the geometric phase close to a QCP \ct{carollo05,zhu06}.
Let us recall the complete ground state given by the product form in 
Eq.~(\ref{prod1}).
To compute the fidelity, let us introduce a family of Hamiltonians generated 
by rotating each spin by an angle $\eta$ about the $z$ direction 
\ct{carollo05,zhu06,patra11}, i.e., $H (\eta) = g_\eta H g_\eta^\dg$ with
$g_\eta = \prod_{\vec n} \exp(i \eta \si^z_{\vec n}/2)$; this unitary 
transformation leaves the energy spectrum in Eq.~(\ref{spectrum}) unaltered. 
Under this rotation, the Pauli spin matrices transform as $\si^x_{\vec n} 
\to \cos \eta ~\si^x_{\vec n} - \sin \eta ~\si^y_{\vec n}$ and 
$\si^y_{\vec n} \to \cos \eta ~\si^y_{\vec n} + \sin \eta ~ 
\si^x_{\vec n}$. Hence the Majorana fermions transform to
\ba a_{\vec n} (\eta) &\equiv& \cos \eta ~a_{\vec n} ~+~ \sin \eta ~
a'_{\vec n}, \non \\
a'_{\vec n} (\eta) &\equiv& \cos \eta ~a'_{\vec n} ~-~ \sin \eta ~a_{\vec n},
\non \\
b_{\vec n} (\eta) &\equiv& \cos \eta ~b_{\vec n} ~-~ \sin \eta ~b'_{\vec n}, 
\non \\
b'_{\vec n} (\eta) &\equiv& \cos \eta ~b'_{\vec n} ~+~ \sin \eta ~b_{\vec n},
\ea
with similar expressions for $a_{\vec k} (\eta)$, $a^\dg_{\vec k} (\eta)$,
etc. The ground state of $H (\eta)$, denoted by $| \Psi (\eta) \rangle$, is 
therefore given by an expression similar to Eq. (\ref{prod1}), with 
$a^\dg_{\vec k}, ~a'^\dg_{\vec k}, ~b^\dg_{\vec k}, ~
b'^\dg_{\vec k}$ being replaced by $a^\dg_{\vec k} (\eta), ~
a'^\dg_{\vec k} (\eta), ~b^\dg_{\vec k} (\eta), ~b'^\dg_{\vec k} (\eta)$.

We now find that the overlap between the ground states for two different
values of $\eta$ is given by
\ba \langle \Psi (\eta_1) | \Psi (\eta_2) \rangle &=& \prod_{\vec k}~ [1 
- \frac{1}{2} ~(1 - \sin \theta_{\vec k}) ~\sin^2 (\eta_1 - \eta_2)], \non \\
{\rm where} ~~\sin \theta_{\vec k} &=& \frac{\beta_{\vec k}}{\sqrt{\al^2_{
\vec k} ~+~ \beta^2_{\vec k}}}. \label{fidel} \ea
(Note that the overlap is unity for both $\eta_2 = \eta_1$ and $\eta_2 = \eta_1
+ \pi$). Eq. (\ref{fidel}) implies that
\ba \ln \langle \Psi (0) | \Psi (d\eta) \rangle &=& - \frac{1}{2} ~(d\eta)^2~
\sum_{\vec k}~ (1 - \sin \theta_{\vec k}) \non \\
&=& - \frac{L}{8A} ~(d\eta)^2~ \int_{\vec k} ~d^2 {\vec k} ~(1 - \sin 
\theta_{\vec k}), \non \\
&& \label{fid_kit} \ea
up to order $(d\eta)^2$, where $A = 4\pi^2 /(3\sqrt{3})$ denotes the area of 
half the Brillouin zone over which the integration is carried out in the 
second equation in (\ref{fid_kit}). (We recall that the number of $\vec k$ 
points in half the Brillouin zone is given by $L/4$). 

The above expression for the fidelity shows that there is no term of 
first order in $d\eta$ in the present case; hence the geometric phase is 
zero. The coefficient of the second order term, $d\eta^2$, yields the 
fidelity susceptibility. Note that this is proportional to $L$ and
does not exhibit any non-analytic behavior as a function of the couplings
$J_1$, $J_2$ and $J_3$. This is not surprising; a rotation of all the
spins is simply given by a unitary transformation, and the system does not
cross a QCP as a result of such a transformation.

\section{Conclusion}

We have studied the ground state fidelity in both the thermodynamic and 
the non-thermodynamic limit for a one-dimensional system of massive Dirac 
fermions with and without interactions 
and in the Kitaev model on the two-dimensional honeycomb lattice. The behavior 
of the fidelity in the one-dimensional Dirac system agrees with the general 
scaling predictions made earlier \ct{rams11}. We have also derived general 
scaling relations for the fidelity close to an AQCP and have verified our 
predictions by using the AQCP present in the Kitaev model. 
Moreover, we observe an additional 
logarithmic correction (in the linear dimension $L$ of the system) in the 
scaling form of the fidelity inside the gapless 
phase of the two-dimensional Kitaev model when $\de L^{1/\nu_{\perp}} \ll 1$. 
Our numerical studies apparently indicates a crossover in scaling around 
$\de L^{1/\nu_{\perp}} \sim 1$. Finally we have considered a rotation of all 
the spins in the Kitaev 
model by an angle $\eta$ about $z$-axis and calculated the fidelity between 
two ground states corresponding to two different values of $\eta$. We have
shown that the geometric phase is absent and the fidelity does not show any 
singularity because no QCP is crossed when such rotations are performed. 

\acknowledgments
AD and VM acknowledge Ayoti Patra for collaboration in related works. 
AD acknowledges CSIR, New Delhi, for financial support and DS acknowledges 
DST, India for Project SR/S2/JCB-44/2010.

\noindent E-mail: $^1$victor.mukherjee@cea.fr \\ 
$^2$dutta@iitk.ac.in \\
$^3$diptiman@cts.iisc.ernet.in

\end{document}